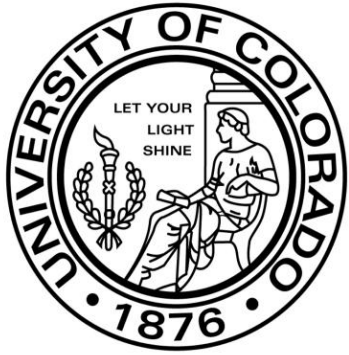

# Intrusions into Privacy in Video Chat Environments: Attacks and Countermeasures

Xinyu Xing[1], Jianxun Dang[2], Richard Han[1], Xue Liu[2], Shivakant Mishra[1]

[1]University of Colorado at Boulder, [2]McGill University

Contact: xingx@colorado.edu

# Intrusions into Privacy in Video Chat Environments: Attacks and Countermeasures

Xinyu Xing[1], Jianxun Dang[2], Richard Han[1], Xue Liu[2], Shivakant Mishra[1]
[1]University of Colorado at Boulder, [2]McGill University
xingx@colorado.edu[1], jianxun.dang@mail.mcgill.ca[2]

**Abstract**

*Video chat systems such as Chatroulette have become increasingly popular as a way to meet and converse one-on-one via video and audio with other users online in an open and interactive manner. At the same time, security and privacy concerns inherent in such communication have been little explored. This paper presents one of the first investigations of the privacy threats found in such video chat systems, identifying three such threats, namely de-anonymization attacks, phishing attacks, and man-in-the-middle attacks. The paper further describes countermeasures against each of these attacks.*

## 1 Introduction

Video chat systems have become increasingly popular, and include such systems as Chatroulette [1], RandomDorm [2], and Omegle [3]. Statistics show that membership in for example Chatroulette has grown by 500 % since 2009. The motivations for using such systems include entertainment, seeking feedback, the desire for companionship, etc. In such systems, individuals are matched with other individuals with whom they can converse via video, audio and text simultaneously. In most such systems, the users do not typically know a priori the other person that they will be matched with for video conversations.

To illuminate the discussion, consider Chatroulette, in which a user will direct their Web browser to www.chatroulette.com to download the Web page for the video chat. The Web page will request the identity of another user from the Chatroulette server, which is then returned in the form of the other user's IP address. "Flash" code embedded within the page then establishes a direct video session with the IP host of the other user, sending local camera and microphone data to the remote user while receiving that remote user's video and audio data. When a user clicks "Next", then the browser will request a new user (their IP address), thus establishing a new video connection. In this example, the server is chiefly involved in supplying IP addresses, and is not involved in relaying any video data. RandomDorm and Omegle exhibit similar architectures.

Privacy has emerged as a concern in these video chat environments. The approximate locations of Chatroulette users can be obtained via a geo-IP lookup, as has been shown at Chatroulettemap.com [4]. Indeed, the founder of Chatroulette states "There is a certain level of anonymity on the Chatroulette that Chatroulette Map takes away" [5]. In addition, conversations that were assumed to be between only a pair of chatters have also shown up unexpectedly on a popular viral video on Youtube, in which a piano player improvises songs based on who he sees on Chatroulette. This video was briefly taken down in part out of privacy concerns [6, 7]. While it is likely unreasonable to expect such systems to perfectly guarantee privacy, it is clear that there is an expectation among the video chat users that such systems will preserve at least some degree of their privacy. Sites like Chatroulette are concerned enough to have begun addressing some of these privacy issues [8].

This paper presents an initial investigation of some additional privacy problems beyond what has been exposed previously in the general class of video chat systems, and identifies three specific classes of attack on such systems while also proposing countermeasures to address these threats. In particular, we find that such systems are subject to enhanced de-anonymization attacks, phishing attacks, and man-in-the-middle attacks. We propose preliminary solutions to deal with these attacks, and observe that there is substantial room for improvement on these techniques. Overall, our belief is that further privacy research is needed to protect this key emerging class of video chat applications. [1]

## 2 Related Work

Video privacy issues in the media space have been well investigated and several defense mechanisms have been enumerated over the past few years. All of them focus on maintaining privacy by using a video blurring method. Over the past decade, video privacy has been usually discussed either in the context of video surveillance systems [9][10][11] or in the context of home-based video conferencing[12][13].

[9] uses a region-based transform-domain scrambling technique to blur the entire video and thus preserve privacy in the context of a video surveillance system. The main drawback of this work is that no clearly observable region remains, i.e. it obscures the whole scene of a video. To address this issue, [10] proposes a respectful cameras system that harnesses statistical learning and classification to obscure faces with solid ellipsoidal overlays and maximize the remaining observable region of the scene. [11] employs a robust face detection and tracking algorithm to obscure human faces in the video. One of the benefits for only blurring an individual's face is that the video surveillance system is still able to monitor what is happening in a specific area.

In the context of home-based video conferencing, researchers have used similar methods to address video privacy issues. To minimize the accompanying loss of privacy, [12][13] describe several image filtering techniques that fil-ter the communicated video streams rather than broadcasting clear video. One purpose to obscure the scene in a video (in the context of home-based video conferencing) is to prevent other video participants from identifying what a participant is doing and what he is wearing. Another objective of using blurring technique in a home-based video conferencing is to make a given scene appropriate for a colleague to view.

---

[1]We have obtained approval from the founder of Chatroulette, Andrey Ternovskiy, to reveal the techniques described in this paper.

Though video blurring tachniques can work appropriately as well as can protect users' privacy in both contexts, they are not appropriate for the new generation of video chat services such as Chatroulette, and in fact, may significantly harm the usability of these new generation of video chat services. This is because the main function of the new generation of video chat services is to bring a user face-to-face via webcam with an interesting person from another corner of the planet. We can imagine that if the services provide users with all blurring faces, the users who use these services will gradually lose their interest in talking to others. We conducted a small-scale experiment using the Chatroulette platform in April 2010. The experimental results have shown that only less than 20% of the users stayed to talk when a user blurred his face.

## 3 Breaching Privacy

Based on our experience with Chatroulette and other similar systems, we outline three attacks against video chat environments.

### 3.1 Enhanced De-anonymization Attack

The most direct attack against video chat environments is to identify users' geographical location. By using a proprietary IP address lookup database and technology, adversaries are able to determine a user geographical location. Chatroulette harnesses Adobe's Stratus platform as a means of reducing their bandwidth costs while providing video services. In general, Chatroulette handles the behind the scenes handshakes involved in making two clients connect, but the actual connection is a direct, peer-to-peer link between the two users. While a connection is established between two peers, audio and video streams encapsulated in UDP streams start to exchange data. An adversary can easily retrieve the source IP address from the header of a packet and use a geo-IP mapping service to find the approximate location of the user. These results have been placed on a Google map at chatroulettemap.com [4].

Anonymity can be further compromised by correlating this location information with other seemingly innocuous information that might be revealed in a casual video chat. For example, after an adversary has determined the approximate location of the other user, the adversary can further engage in some small talk. One of the most common conversations that people have with each other on chatroulette is "How are you doing today?" followed by "what's your name?". Answering the questions correctly implies that one reveals one's private information.

Correlating these answers with location can very quickly allow an adversary to converge on the user's identity, even though the user thought that revealing just their first name would keep their identity secret. For example, adversaries can easily search on Facebook by a user's first name *and* the physical location (searches can be further refined based on school and workplace). Although the query issued by the adversary might return tens (even several hundreds) of results related to the name and general physical location (the number of returned results depends on the uniqueness of the name, that is, Rick is more common than Xinyu on Facebook), the adversary can easily and quickly filter the personal profiles by comparing the appearance on screen with the thumbnail pictures shown in one's social profile.

To validate our enhanced de-anonymization attack, we conducted an experiment over the Chatroulette platform. To do this experiment, we implemented software to retrieve the incoming and outgoing UDP packets. A screenshot of this software has been provided in Figure 1. Based on the packets that the software collects, the software displays the top-five UDP communications (here we rank UDP communications based on the volume of UDP packets) to distinguish which UDP traffic is associated with the Chatroulette application. Notice that other applications running on the machine may also generate UDP traffic but the volume of the traffic is usually lower than the volume of traffic generated by other applications. Furthermore, our software utilizes the GeoIP library to convert an IP address to a physical address. We verified that it was possible to apply reverse geo-location mapping to obtain a remote chatter's geographical location, e.g. Denver, Colorado. We could then search Facebook using the first name provided along with the location to retrieve a narrowed down search of a couple hundred names. We found this list could be manually but quickly searched in real time by comparing photos to uncover the identity of the person while the conversation was ongoing. The user's identity could be revealed before the conversation had finished. Thus, an individual who believes that they are sufficiently anonymized by just revealing their first name was not safe from having their identity revealed using this enhanced de-anonymization attack.

### 3.2 Phishing Attack

In this type of attack, an adversary poses as an attractive person to solicit potentially sensitive information from other chat users, such as their name, Skype account, Facebook account, telephone numbers, etc. Unlike traditional phishing attacks, which are typically carried out by email or by hijacking instant messages[14], an adversary in a video chat system can simply play a video that he prepared in advance to lure unsuspecting individuals into a conversation with the adversary, potentially divulging sensitive information. In this type of attack, an adversary replaces their own video with that of an attractive young lady or a gentleman. For example, prior work over the Chatroulette platform has shown that young ladies are more likely to be accosted by Chatroulette male users [15]. The adversary shows a largely head-and-shoulders shot that appears plausibly to be interacting every now and then with the screen, perhaps typing in something on occasion. The audio is disabled, and the adversary claims this is either because he is working from a desktop, which typically doesn't have a built-in microphone, or they cannot speak English well and would prefer text messages. Thus, the primary interaction is with text messages through the chat system. In our tests, most of the users disabled audio, and no users asked us to reenable our disabled audio.

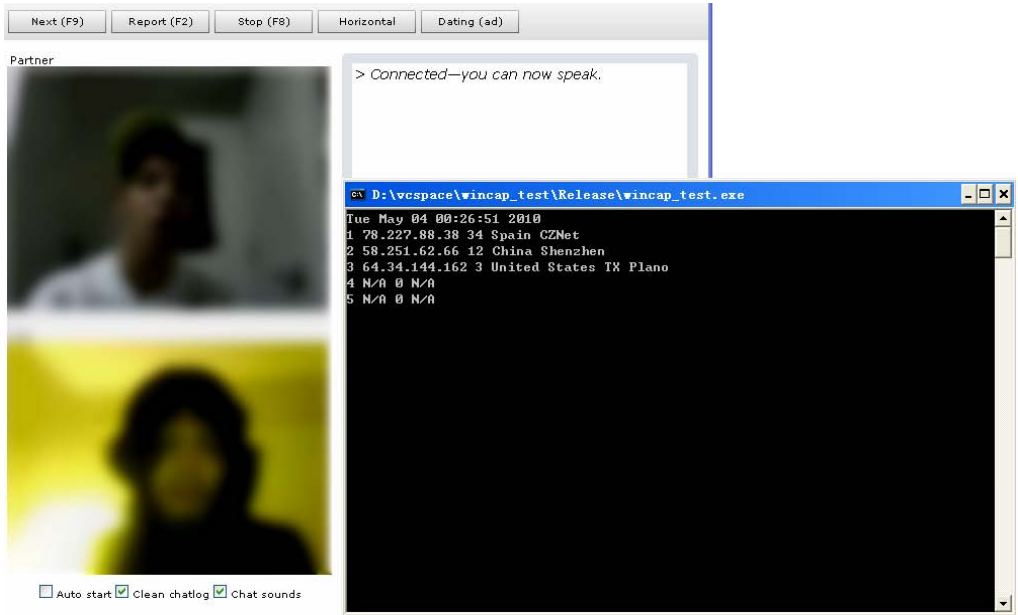

**Figure 1. A screenshot of the software which has been used in deanonymization attacks.**

As a result, the remote user is deceived into believing they are interacting with an attractive approachable person. During the conversation, the adversary attempts to steer a polite conversation, gains the user's trust and thus can seek to extract a user's private information, such as name, Facebook account, Skype account and even telephone number, etc. For example, the adversary can attempt to steer a casual conversation, bring sexual innuendos into a conversation and use a screen recording program to record the conversation. Furthermore, the adversary can utilize the victim's private information to search for the victim's social relationships (e.g. using the victim's Facebook account to identify his friends). This information can be used to blackmail the victim, e.g. the adversary may threaten to broadcast the recorded conversation to the victim's friends or family. Even worse, the true attacker can remain anonymous hidden behind their phishing personae.

To investigate the feasibility of the phishing attack, we designed an experiment in which we substituted a fake video personae in place of the normal webcam. In our experiment, we first recorded a video in which a lady sitting in front of the webcam pretends to chat to a video chat user. Then, we redirected the video chat software to obtain the video from the recorded video stream, rather than the webcam. For example, in Chatroulette, the Flash software asks you which webcam you'd like to use, and there is an option for selecting a virtual webcam. One only needs to play the fake video in a separate window, capture and transfer this video as the source of the virtual webcam. The resulting software is shown in Figure 2.

To further disguise the adversary, we emulated a low-bandwidth network. We edited the video that we prepared by changing the frame frequency. To do this, we select partial frames to form our video. In our experiment, we only retain those frames which happens every 2 seconds. For example, a raw 6-second video contains 144 frames. We only retain the 1st, 48th, 96th and 144th frame. Intuitively, a Chatroulette user is not able to see a continuous video stream and thus he may think that the connection with the adversary is experiencing a slow speed. Another reason why we slow down our video is that an adversary can prevent a Chatroulette user from asking the adversary to verify that the video is real (e.g. a Chatroulette user may ask the adversary to raise her hand). Since the video is not continuous, the adversary can easily make an excuse – the motion of raising hand may not be able to be transmitted to the Chatroulette user. In practice, one of our observations over a one-hour experiment is that only one out of fifteen chatroulette users asked us to verify the reality of the video. And after we made the excuse to explain the reason why he cannot see our rising hand, he simply believed our excuse. This confirms that it is possible to deceive unsuspecting users, and even one more wary user, as to whom they are conversing with by substituting a poor quality video.

Though we did not proceed further to extract personal information, we believe that once an adversary has gained the other user's trust, especially with a photogenic presence, then psychologically it becomes much easier to ask them to divulge some personal information, say asking to friend them on Facebook, or contact them via Skype. In that respect, the bar is lowered for using the obtained information for nefarious purposes.

### 3.3 Man-in-the-Middle Attack

Our observation is that video chat users can become vulnerable to man-in-the-middle (MIM) attacks. While two video chat users may believe that they are talking directly to one another, it is possible to insert a man-in-the-middle who can eavesdrop on their conversation without the two end users being aware that their conversation is being ob-served and/or recorded. Combined with de-anonymization attacks, then it is possible to compromise a user's identity and threaten such observed/recorded users with blackmail.

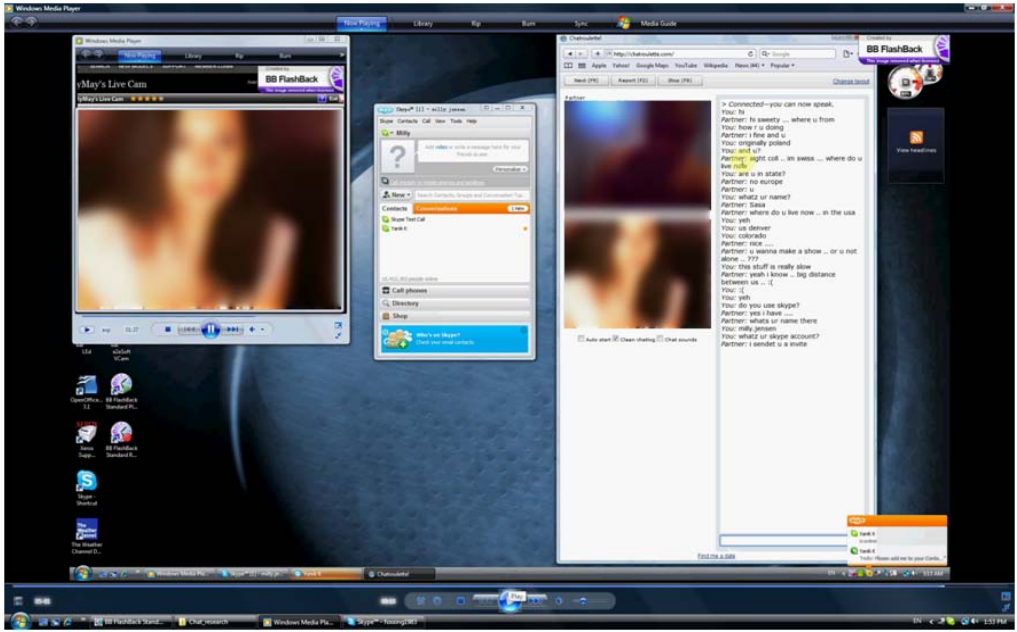

**Figure 2. Using a video to carry out a phishing attack.**

An MIM attack would work as follows: an adversary could open up two video chat windows, one talking with remote User 1, and another talking with remote User 2. Then, using a similar mechanism to the phishing attack, the adversary could substitute User 2's video stream in place of its own webcam stream to User 1. Similarly, the adversary could substitute User 1's stream in place of its own stream to User 2. By crossing the streams, the adversary can observe the dialogue between Users 1 and 2 without either of the two users realizing their conversation is being observed.

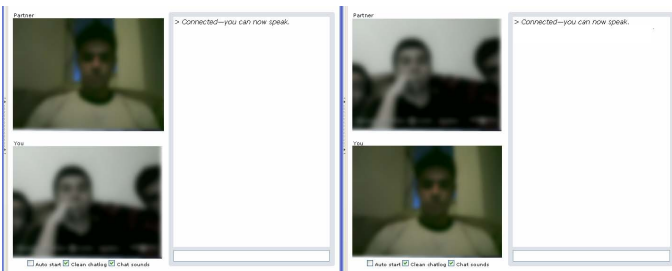

**Figure 3. A video relay based man-in-the-middle attack.**

We term this type of MIM attack a video relay (VR) attack. Figure 3 illustrates the process of this VR-based MIM attack. An adversary uses a virtual camera program to relay victims' video streams to each other. In this case, the adversary relays the victim (the person shown in the top-left window) Alice's video stream to the victim Bob (the person shown in the top-right window) by using a virtual camera program. Similarly, Bob's video stream is redirected to Alice. Notice that a VR-based MIM attack requires the adversary to utilize two different virtual camera programs for the reason that one of the victims will see his own video stream if only one virtual camera program is used. To illustrate this, let's take a look at an example. Suppose that an adversary uses a single virtual camera program to redirect Alice's video stream to Bob. Without doubt, Bob is able to see Alice's video. However, since the connection session between Alice and the adversary also uses the same virtual camera program, Alice will see her own video showing in the partner window.

We believe Chatroulette as currently implemented is vulnerable to such MIM attacks. An adversary is able to connect to two or more Chatroulette users simultaneously. By redirecting conversation content (including video, audio and text content) to another Chatroulette user, the adversary is able to invisibly eavesdrop on the conversation between two victims.

In addition to the VR-based MIM attack, a more general packet relay (PR) attack is possible. For example, Chatroulette uses the Real-Time Media Flow Protocol (RTMFP), which is encapsulated in a UDP message, which indicates that the video and audio messages exchanged between a pair of Chatroulette users are encrypted. In order to achieve PR-based MIM attacks, an adversary therefore cannot simply relay the messages that he received from one victim to another victim because each pair of Chatroulette users share different keys. Instead, the adversary has to first decrypt the message exchanged between him and one victim, use the key shared with another victim to encrypt the message and then send the new encrypted message to the other victim. Thus, even if Chatroulette used https with TLS/SSL to encrypt both video sessions, the MIM can decrypt an arriving stream from a user and reencrypt the data as an output stream to the other user.

Generally, the PR-based MIM attack is more holistic than VR-based attacks for the reason that the PR-based attack can also relay victim text and audio messages while the VR-based attack has to disable the audio devices and even may involve human interruption, e.g. manually redirecting victims' text messages.

### 3.4 Discussion

*1. In a video chat system, users (knowingly) reveal video of their faces. Do these users have any expectation of security or privacy while using such systems?*

In any social networking system, or in fact any Internet service, users are generally well-aware of the fact that some of their private information will be revealed. However, there is a general belief that only the information that a user (explicitly) discloses via these systems is revealed and it is revealed to only to those users that this user authorizes. Indeed, systems such as Facebook and Skype attempt to provide system support for such security and privacy. We believe that the users of video chat systems such as Chatroulette have such expectations. For example, it is reasonable to assume that a Chatroulette user believes that he/she is chatting with a real user, or no one else other than the user at other end is seeing/receiving his/her information.

*2. With today's Web, including face matching technology, is it reasonable to expect a video chat service such as Chatroulette to still have privacy?*

Most current social networking services distinguish between public and private profiles of a user. A public profile is generally available to all users, while access to the private profile is restricted. Chatroulette is similar. Video of an online user's face is available to all users, but contents exchanged after a video chat session has been established must be authenticated, secure and private. Indeed, correlating public information available from one website (e.g. face video in Chatroulette) with information available from other websites can be done and this may reveal additional information about a user. However, this is a general problem facing all Internet services that care about privacy, and is not unique to Chatroulette.

*3. If a video chat service is anonymous for both sides, why does it matter that eavesdropping using man-in-the-middle attack is feasible?*

A key advantage an attacker gains by launching the above-mentioned man-in-the-middle attack is anonymity. The two users that are led to believe that they are chatting directly have no way to determine the identity of the attacker.

*4. How do we deal with the audio stream apart from the video stream?*

All attacks, apart from the PR-based MIM attack, require an adversary to disable audio devices. Unlike a laptop in which a microphone is usually embedded, it is quite common that a desktop is not equipped with a built-in microphone. Therefore, disabling the audio input device (i.e. microphone) does not influence the usability of Chatroulette. Furthermore, based on our observations, more than half of Chatroulette users either disable or are not equipped with audio input devices. Therefore, all attacks described above are not easily detected when the adversary disables his audio input devices.

*5. How do we get Chatroulette users' attention?*

According to a previous study [15] on Chatroulette, 71% of its users were male. Female Chatroulette users usually get more attention from male users. Furthermore, we conducted an experiment by using a video of an attractive lady sitting in front of the webcam. Based on our observation, over 95% of male Chatroulette users passively started a conversation with us when they saw the video. Therefore, our experiments introduced above take advantage of this human behavior to verify the possibility of the proposed attacks.

*6. How do we use online social networks to search for Chatroulette users' profile?*

There are several searching mechanisms on online social networks that can help an adversary de-anonymize victim's privacy. One of the most directed mechanisms is to use the searching functions that an online social network provides. For example, Facebook allows users to perform friend search. Based on the returned results, Facebook also offers a function that could help users to refine the returned results by using location information (see Figure 4). In addition, if the number of the returned results are still in a high range, the adversary can also use a victim's personal interests to match the victim's profile and thus filter their searching results if the victim reveals his hobbies during the conversation.

## 4 Countermeasures

To address these security attacks, we provide some security suggestions and possible countermeasures.

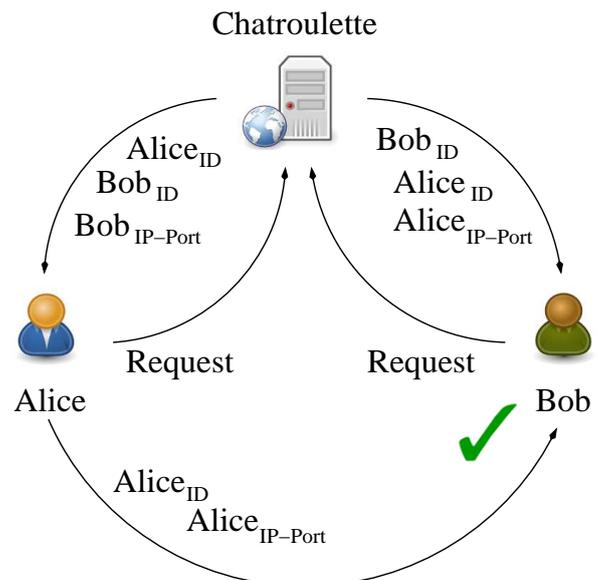

**Figure 5. Operations in Chatroulette.**

### 4.1 De-anonymization Countermeasures

As introduced in Section 3.1, the de-anonymization attack mainly uses the victim's IP address along with victim's name and video to carry out a malicious attack. The most intuitive defense against de-anonymization attacks is to blur video chat users' faces and thus protect their identities. However, this may significantly harm the applicability of the face-to-face online conversation service. In addition, the computation-intensive technique for blurring chat users' face may not always work properly. Consequently, another simple solution which comes to our mind is to hide users' IP addresses and suggest a video chat user to use an alias instead of his actual name before getting into a conversation with a stranger. In general, the simplest mechanism to anonymize one's IP address is to harness anonymous services (e.g. Tor[16]).

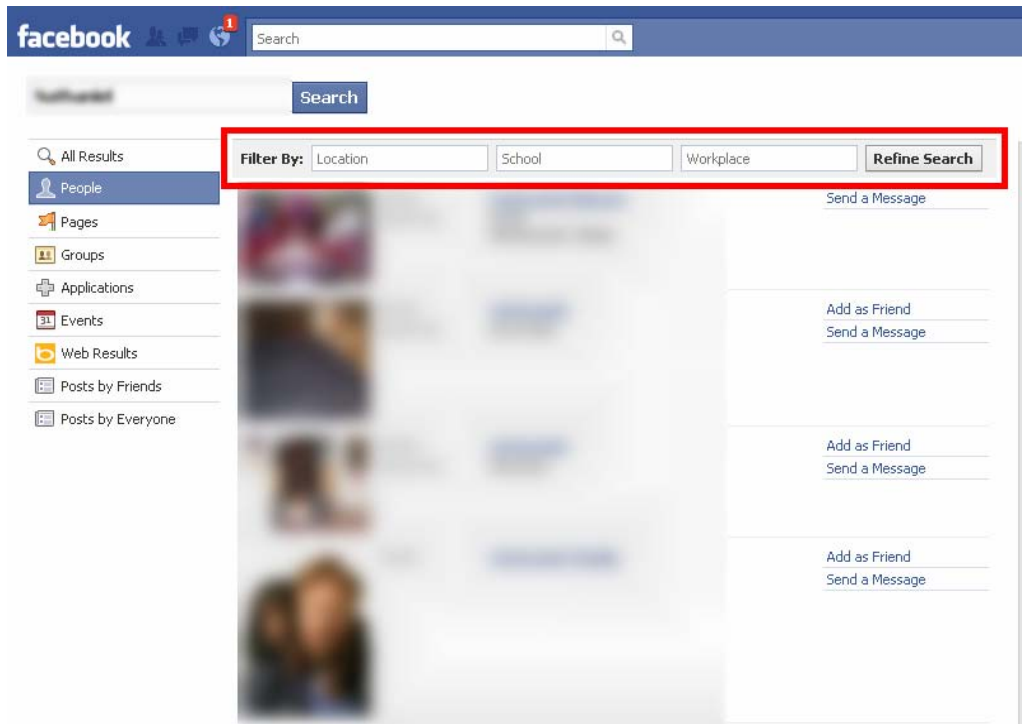

**Figure 4. Search and refine victims by using Facebook.**

According to our experiment, we however found out that anonymous services such as Tor do not work when combined with a service like Chatroulette. As shown in Figure 5, Alice and Bob send their request for a new chat session (including IP address and Port) to Chatroulette. Chatroulette uses Alice's (and Bob's) IP address and port to generate user IDs for Alice and Bob respectively. Chatroulette's server then sends to Bob the four-tuple (Bob's ID, Alice's ID, Alice's IP, Alice's UDP port), and to Alice the four-tuple (Alice's ID, Bob's ID, Bob's IP, Bob's port). Bob now has enough information about Alice to verify that a request for a video chat did indeed come from Alice, by checking the request matches Alice's ID, IP, and port number. This filters out spurious requests for video chats.

However, when Tor software is mounted in any Chatroulette user, Chatroulette's service cannot work properly. Figure 6 provides a reasonable explanation for why using Tor doesn't quite work. Assume that Alice installed a Tor software on her machine and uses the software to establish a connection with Chatroulette and Bob. When Alice registers herself with Chatroulette, the IP address and port number seen by Chatroulette is from the TOR network. This information is then communicated to Bob as Alice's credentials. When Alice sends a request to Bob initiate a chat with him, her packet will include a different IP address and port number, i.e. if she uses TOR, then her packet's source IP will contain a different IP and presumably port assigned by a different TOR proxy. Even if she avoids TOR in talking directly with BOB, namely uses her own IP and port, there will still fail to be a match at Bob's. In either case, the IP address and port provide by Alice fail to match the ones that Bob has on record. Therefore, the connection fails.

One possible solution to this problem is to use the same proxy entity to establish a connection with the Chatroulette service and the partner. While using the same proxy to establish connection, the Chatroulette service can generate a user ID for a Chatroulette user by using the proxy IP address and port. Similarly, when the Chatroulette user connects to another user (through the same proxy) that Chatroulette assigns, the connection between these two users can work properly because the IP address and port of the proxy is compatible with the User ID that Chatroulette assigned. Here the proxy has to be trusted by two peers. Although the trusted proxy may perfectly hide a user's IP address and port, there is still a potential performance limitation. The benefit of Adobe's Stratus is to help Chatroulette reduce bandwidth costs. Therefore, the involvement of a fixed proxy may sacrifice Chatroulette users' performance in terms of bandwidth and latency. In addition, we would also suggest a Chatroulette user to use an alias to substitute his actual name, since an alias can make malicious de-anonymization more difficult.

### 4.2 Phishing Countermeasures

Intuitively, phishing attacks can be easily detected and defended against. For instance, a Chatroulette user could ask his partner to prove that the video stream he is watching is real by asking his partner to raise his hand or wink his eyes etc. However, an adversary can also make an excuse such as "the network is slow. I raised my hand but you did not see it ..." and thus evade this check. In this case, the Chatroulette user should ask his partner to maintain a gesture (such as raising left hand) for a period of time until the frame with the gesture is transmitted back to the Chatroulette user. In addition, another intuitive method to detect fake video stream, though it might not be that efficient and effective, is to check with audio stream. However, the adversary may have a plausible reason for disabling his audio input devices. Therefore, we suggest that when meeting a stranger with audio disabled, the Chatroulette user should verify the authenticity of the video stream that he watches by requesting a long-lasting gesture be repeated.

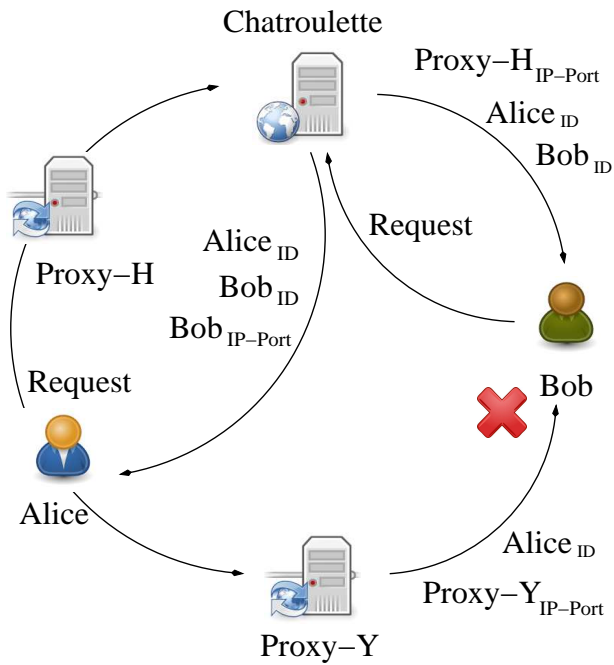

**Figure 6. Operation of Tor in Chatroulette.**

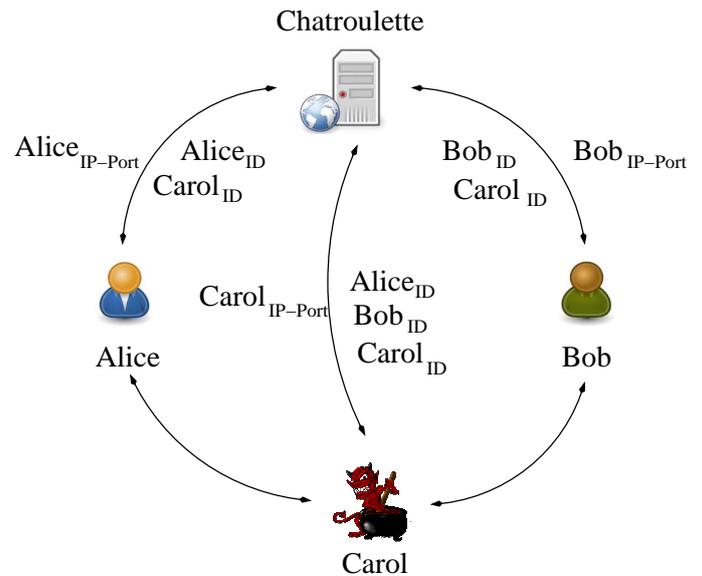

**Figure 7. Man-in-the-Middle Attacks.**

However, a recent movie – *Avatar* – throws new light on an advanced phishing attack. Inspired by that movie in which scientists develop an Avatar program which enables people to control their own Avatar, an adversary could develop a program that would control the motion of a realistic-looking avatar shown in a video stream by either typing on a keyboard or clicking a mouse. In this sense, the attacker's goal is to present a fake personae that can pass a video version of the Turing test [17]. In practice, there have already been several game programs that have implemented similar functions [18]. Notice that the game program uses 2-3D technology to describe the object in the video rather than the real people.

Another mechanism for preventing phishing attacks is to identify whether the Chatroulette user that you get a conversation with is using a virtual webcam rather than an actual webcam. Since the phishing attacks are carried out by using a virtual webcam to fool Chatroulette users, terminating or identifying the virtual webcam means terminating phishing attacks. A patch of Tencent QQ[19] has been implemented to identify virtual webcams in Tencent's instant messaging service platform. The basic idea of this technique is to blacklist all mainstream virtual webcam software. When a device's driver (either an actual device or a virtual device) is installed on Windows system, the driver will be automatically registered at the Windows system registry. Since the instant message software has already blacklisted all the virtual webcams, the instant message software will notice the use of a virtual webcam to the other user when a virtual webcam is being used. However, the main drawback of this technique is inherent from the fact that the Windows system allows its users to modify their registry. Therefore, an adversary can easily bypass its detection by changing the virtual webcam driver's key value in the Windows system registry into one that is not shown in the blacklist. The best solution that we suggest in this paper is to develop a program that could automatically lock the actual webcam device and disable the virtual webcam. At present, we are in the middle of the software design to distinguish virtual webcams from the actual webcams.

### 4.3 Man-in-the-Middle Countermeasures

In the current Chatroulette service, which does not support anonymous IP addresses, one approach to address MIM attacks is to seek to verify IP addresses between one another. Take an example of Alice and Bob. As indicated in Figure 7, the conversation content between Alice and Bob is relayed by a malicious user – Carol. For both Alice and Bob, the connection peer is Carol. Therefore, Alice and Bob are able to realize that the video streams transmitted in between all come from Carol's IP address. If Alice and Bob are able to exchange the IP address of their incoming video streams, they could identify that a malicious attack is happening in between because the IP address that they see is exactly the same. However, exchanging IP address through the mali-cious user is difficult for the reason that Carol can modify the content of the packets that she receives. One solution is to watermark the IP address of Alice and Bob's incoming video streams into their outgoing video streams. In this way, Bob would receive video watermarked with Alice's IP source address, and compare that with the IP address of the source. If the two do not match, then Bob can suspect a MIM. The MIM adversary is thus forced to do compute-intensive processing on the video stream to extract the existing watermark and replace it with the MIM's IP watermark. This raises the difficulty level on an adversary, though a well-equipped adversary could still overcome such a defense.

Recall the protection mechanism that we introduced in Section 4.1. The way that we protect de-anonymization attacks is to use a proxy entity. Therefore, there is a possibility that the protection solution to de-anonymization attacks can be identified as a MIM attack if Alice and Bob are using the same proxy to hide their IP address. To address this problem, a possible solution is to whitelist all the third-party trusted proxies at the Chatroulette website. Every time, when a Chatroulette user asks Chatroulette to assign him a partner, Chatroulette also sends the user the list of those trusted third-party proxies. So, when the watermarking is the same, Alice and Bob can first check whether the incoming IP address is different from the one that has been whitelisted before terminating the conversation.

## 5 Conclusion

Although video chat systems such as Chatroulette and RandomDorm have received great attention, we have demonstrated that current security and privacy issues of these systems have been neglected. According to our experiments over the Chatroulette platform, we investigated some potential security and privacy vulnerabilities including de-anonymization attacks, phishing attacks and man-in-the-middle attacks and presented the corresponding countermeasures. IP anonymization with Tor was proposed as a countermeasure to de-anonymization attacks, but was found to introduce some compatibility problems with a service like Chatroulette. Phishing attacks could be counteracted by requiring visible authentication, e.g. raise your right hand, but a sophisticated adversary could thwart this attack with an avatar. Virtual webcam detection was offered as an option, but an adversary could thwart this approach as well. MIM attacks could be addressed by watermarking videos with the IP address of their destination.

We have just begun to scratch the surface of interesting attacks on and countermeasures for video chat systems, and we hope that this paper will stimulate further discussion on many new approaches to protecting privacy in video chat systems. For example, our countermeasures for anonymization have weaknesses, as do our solutions for phishing attacks. We have begun to consider cryptographic approaches that bind more strongly a video chatter's source address to the video source. We will probe these problems more deeply as our future work progresses.

## 6 References

[1] "Chatroulette," http://www.chatroulette.com.

[2] "Randomdorm (for college students)," http://randomdorm.com/emails/new.

[3] "Omegle," http://omegle.com/.

[4] "Chatroulettemap.com," http://chatroulettemap.com.

[5] "One on one: Andrey ternovskiy, creator of chatroulette. nytimes." http://bits.blogs.nytimes.com/2010/03/12/one-on-one-andrey-ternovskiy-creator-of-chatroulette/?src=twt&twt=nytimesbits.

[6] "Chatroulette improv piano player removed from youtube." http://mashable.com/2010/03/22/merton-removed-youtube/.

[7] "Chatroulette piano improv guy merton removed from youtube." http://boingboing.net/2010/03/23/chatroulette-piano-i.html.

[8] "Chatroulette founder working to preserve user privacy," http://site14.fourfiveone.com/2010/03/13/chatroulette-founder-working-to-preserve-user-privacy-2/.

[9] F. Dufaux, M. Ouaret, Y. Abdeljaoued, A. Navarro, F. Vergnenegre, and T. Ebrahimi, "Privacy enabling technology for video surveillance," in *Proc. of SPIE Mobile Multimedia/Image Processing for Military and Security Applications*, 2006.

[10] J. Schiff, M. Meingast, D. Mulligan, S. S. Sastry, and K. Goldberg, "Respectful cameras: detecting visual markers in real-time to address privacy concerns," in *Proc. of Intelligent Robots and Systems*, 2007.

[11] D. Chen, Y. Chang, R. Yan, and J. Yang, "Tools for protecting the privacy of specific individuals in video," in *EURASIP Journal on Applied Signal Processing*, 2007.

[12] Q. A. Zhao and J. T. Stasko, "Evaluating image filtering based techniques in media space applications," in *Proc. of the 1998 ACM conference on Computer supported cooperative work*, 1998.

[13] C. NEUSTAEDTER, S. GREENBERG, and M. BOYLE, "Blur filtration fails to preserve privacy for home-based video conferencing," in *ACM Transactions on Computer-Human Interaction*, 2006.

[14] N. Chou, R. Ledesma, Y. Teraguchi, and J. C. Mitchell, "Client-side defense against web-based identity theft," in *Proc. of NDSS*, 2004.

[15] "The awl," http://www.theawl.com/2010/02/chatroulette-explained-hot-girls-dont-get-nexted.

[16] R. Dingledine, N. Mathewson, and P. Syverson, "Tor: the second-generation onion router," in *Proc. of the 13th conference on USENIX Security Symposium*, 2004.

[17] "Turing test." http://en.wikipedia.org/wiki/Turing_test.

[18] "Gamespy," http://planethalflife.gamespy.com/.

[19] "Tencent," http://www.tencent.com/en-us/index.shtml.